\documentclass[doublecol]{epl2} 

\usepackage{graphicx}
\usepackage{amsmath}
\usepackage{amsfonts}
\usepackage{amssymb}

\newcommand{\beqs}{\begin{eqnarray*}}
\newcommand{\eeqs}{\end{eqnarray*}}
\newcommand{\beq}{\begin{eqnarray}}
\newcommand{\eeq}{\end{eqnarray}}

\newcommand{\aS}{|S|}
\newcommand{\bit}{\begin{itemize}}
\newcommand{\eit}{\end{itemize}}

\newcommand{\al}{\alpha}
\newcommand{\p}{\partial}
\newcommand{\Voa}{V_{0,\al}}
\newcommand{\Woa}{W_{0,\al}}

\title{Dewetting of thin polymer films: Influence of interface evolution}

\author{Falko Ziebert \and Elie Rapha\"el}
\shortauthor{F. Ziebert, E. Rapha\"el}

\institute{                    
  Laboratoire de Physico-Chimie Th\'eorique - UMR CNRS Gulliver 7083, ESPCI, 10 rue Vauquelin, F-75231 Paris, France
}
\abstract{The dewetting dynamics of ultrathin polymer films, 
e.g. in the model system of polystyrene 
on a polydimethylsiloxane-covered substrate, exhibits interesting behavior
like a fast decay of the dewetting velocity and a maximum in the width of
the built-up rim in the course of time. These features have been recently ascribed to 
the relaxation of residual stresses in the film that stem from 
the nonequilibrium preparation of the samples.
Recent experiments by Copp\'ee et al. on PS with low molecular weight, where such stresses could
not be evidenced, showed however similar behavior. 
By scaling arguments and numerical solution of a thin film viscoelastic model 
we show that the maximum in the width of the rim can be caused 
by a temporal evolution of the friction coefficient (or equivalently of the slip length), 
for which  we discuss two possible mechanisms. 
In addition, the maximum in the width is affected by the sample age.
As a consequence,
knowing the temporal behavior of friction (or slip length)
in principle allows to measure the aging dynamics of a polymer-polymer interface 
by simple dewetting experiments.
}

\begin{document}

\maketitle

\section{Introduction}

Dewetting experiments \cite{Reiter:2001,Herminghaus:2001,Reiter:2003.1} have been  
proven to be a simple tool to investigate thin polymer films  
on various substrates \cite{Bucknall:2004,Blosseyreview}. 
Especially the model system of the incompatible 
polymers polystyrene (PS) on polydimethylsiloxane (PDMS) grafted on a silicium substrate
has attracted numerous experimental investigations. 
There is consensus that friction and slippage at the interface, as well as
the viscoelasticity of the dewetting film are crucial to 
understand the experiments \cite{Herminghaus:2003,Vilmin_slippery}. 
To account for the occurrence of a maximum in the width of the rim in the course of time, 
and to interpret the effects of sample aging 
on this peculiar feature of the thin films' dewetting dynamics, 
residual stresses in the PS film have 
been shown to be important \cite{Reiter:2005,Vilmin_nlfric,Reiter:2007.2}.
These stresses are supposed to stem from the sample preparation process (usually spin-coating) 
and represent an additional degree of freedom that can (partially) relax during dewetting.
For high enough molecular weights, residual stresses have been
clearly evidenced in various systems \cite{Reiter:2005,Yang2006,Fretigny}. 
However, their relaxation well below the glass temperature, needed
to interpret the effects of aging \cite{Reiter:2005}, is currently debated.
An additional puzzle is the fact that for films of low molecular weight 
no such relaxation of stress could be evidenced, while a maximum in the rim width 
and its dependence on aging have still been found.
Very recently, in parallel to dewetting experiments the interface between the 
two polymer species has been studied by neutron reflectometry \cite{Pascal_new}
and interdiffusion of the PS-PDMS interface has been found, 
for samples aged well below the PS bulk glass temperature.

In this letter we show by scaling arguments and demonstrate by numerical solution 
of a thin film model that a slow increase in 
the friction coefficient or, equivalently, a decrease of the slip length 
can give rise to a maximum in the rim width. 
This evolution of friction/slip
might be due to several mechanisms: first, as suggested by the neutron reflectometry 
measurements \cite{Pascal_new}, a slow roughening of the film-substrate (PS-PDMS)
interface with concomitant increase in friction may be responsible. 
A second possibility would be that (for the lowest PDMS grafting densities)
a small number of PS chains may enter the PDMS brush and become attached to the substrate.
It is known that minute amounts of such 'connectors' \cite{EliePGG,ReiterEPL,ReiterKhanna} 
already decrease the slip length. 

A second finding is that the value of the maximum rim width decreases monotonously 
with the friction coefficient at the beginning of the dewetting process. 
This could explain the experimentally observed decrease of the
maximum with the sample age.
Thus knowing the evolution of the friction coefficient allows in principle to relate 
the maximum width to the aging history.
As studied in Refs.~\cite{Reiter:2005,Vilmin_nlfric,Raphael:2006.1,FZER1},
a similar connection between rim width and aging time 
can also be caused by relaxing internal residual stresses. 
One would expect that stress relaxation is the dominating mechanism for high molecular weight films,
while for low molecular weight films interface changes are responsible. 
In general, the two nontrivial processes, stress relaxation in bulk and evolution of the interface, 
might be present at a time, which leaves interpretation of experimental results an intricate task.  
The specific dynamics of the PDMS brush-PS melt interface being unknown, 
drawing conclusions from the dependence of the rim width
on the sample aging, as suggested above, is only a rough estimate at present.
Dynamic self-consistent field techniques or molecular simulations
would be needed to establish this dynamics.

\section{Which processes can lead to a maximum in the rim width ? A scaling argument}

First we want to give a simple argument, why a slow evolution of the interface
makes the occurrence of a maximum in the rim width possible.
For dewetting of a viscoelastic film, scaling arguments have been used
to qualitatively understand the dynamics \cite{FBW:1997,Raphael:2006.1}. 
The balance of the work done by the driving force 
and the dissipation by friction 
reads approximately
$|S|V(t)\simeq\zeta W(t) V(t)^2$ or
\beq\label{en_bal}
\frac{|S|}{\zeta} \simeq W(t) V(t)\,.
\eeq
Here $V(t)$ is the velocity at the dewetting edge and $W(t)$ is the width of the rim.
The driving force acting on the rim is the negative of
the spreading parameter $S=\gamma_{sv}-\gamma_{sl}-\gamma$
(where the $\gamma$'s are the interface energies 
for the substrate-vapor, substrate-liquid and liquid-vapor
interfaces, respectively), and 
$\zeta$ is the friction coefficient associated to the polymer-polymer interface.
In general, the friction will be nonlinear (see below), but for the simple argument given here
this is not of importance.
In addition to this balance, one needs the fact that
the dewetting velocity is monotonously decreasing -
there is no mechanism, and also no experimental evidence,
for a speed-up of the dewetting process. 

What possibilities does Eq.~(\ref{en_bal})
allow for $W(t)$ to have a maximum?
Clearly, if $W$ has passed through a maximum, 
it has to decrease for some time
{\it simultaneously} with the monotonously decreasing velocity. 
This is not possible for Eq.~(\ref{en_bal}) if
the left hand side is a constant. A {\it necessary condition} for a maximum in the width is thus
that the effective driving force $|S|/\zeta$ decreases in the course of time.
There are several possibilities to achieve such a decrease:
The first one is motivated by residual stresses stemming from spin-coating. 
Introduction of a residual stress $\sigma(t)$, that is allowed to relax in the course of time,
leads to a renormalization of the driving force $\aS\rightarrow\aS+h_0\sigma(t)$ 
\cite{Raphael:2006.1,FZER1} (where $h_0$ is the film thickness). 
Accordingly, upon stress relaxation the driving force decreases.

The second possibility, suggested by the interface evolution evidenced in Ref.~\cite{Pascal_new},
is an increase in the friction coefficient $\zeta\rightarrow\zeta(t)$ due to 
roughening of the liquid-substrate interface. 
Additional to the increase in friction, interface roughening also leads to a decrease of the
driving force $|S|$, as discussed for autophobic dewetting in \cite{Kerle_Klein}, and can
even lead to a cross-over to wetting in that case.
For the incompatible system PS-PDMS, where the
PS-air surface energy is dominating, this latter effect is however not probable
for the expected microscopically small roughnesses.

A third possibility, also associated to tiny changes in the interface,
is the successive attachment of a few PS chains to the substrate where
the PDMS-coverage is low or defective.
This process leads to a decrease in the driving force $|S|\rightarrow|S|-\nu(t) l f$, where
$\nu$ is the areal density of such 'connector' chains, $l$ the length they are stretched
upon pull-out and $f$ the pull-out force \cite{EliePGG,ReiterEPL}.
Additionally the friction increases due to the connectors 
(in simplest approximation $\zeta\rightarrow\zeta+\nu(t)\kappa$
with $\kappa$ the friction coefficient of connectors; 
for a more accurate treatment, see \cite{Ajd_FBW_PGG}).

In the following we would like to exemplify the second case further.
We first give an estimate of how the interface evolution 
gives rise to an increasing friction coefficient. Later on
this is implemented in a thin film model and its influence
on the dewetting is studied.

\section{Interface evolution}

Polymer-polymer interfaces are well studied \cite{Jonesbook,Budkowski}, 
although mostly at equilibrium.
The equilibrium width of an immiscible interface 
is given by  $\delta\propto a/\sqrt{\chi}$ \cite{Helfand},
where $a$ is the Kuhn length and $\chi$ is the Flory interaction parameter.
A simple scaling argument \cite{DeGennesCRAD90}, consistent also with 
a full statistical treatment \cite{FredrEPJB},
states that the slip length $b$ at the interface is given by 
the ratio of bulk and interfacial viscosity and scales as
$b\propto N/\delta^2$.
The factor $N$ is the degree of polymerization and leads to the well known 
large slip lengths for polymers \cite{DeGennesCRAD79}.
Alternatively, the slippage length can be defined as $b=\eta/\zeta$,
implying that the friction coefficient scales like $\zeta\propto\delta^2$.
This scaling should hold for the interface between two immiscible melts,
for a brush-melt interface the situation is slightly different.
The structure of brushes in various matrix polymers has been
studied in Ref.~\cite{Clarke_brush}, and the width for the immiscible case
has been found to be narrow, but larger than the width between melts.
The effect of the grafting density on the interface width is unclear, but
one would expect that the grafting density of 
the PDMS-brush should not be too high to 
prevent total stretching and allow for mobility of the brush chains.
Indeed, as reported in Ref.~\cite{Pascal_new}, roughening was
pronounced only for the lowest grafting densities where the
interface width increased from 1 to 4 nm  in the course of time.
It should also be mentioned that polymer interfaces display 
thermally activated capillary waves \cite{SferrazzaPRL}, 
which lead to a higher apparent interface width.

As far as the dynamics is concerned, 
the evolution of the interface towards its equilibrium value due to
chain interdiffusion and its influence on the dewetting via 
the slip length or friction is difficult to determine. 
Strictly speaking it requires to solve a hydrodynamic problem 
for a system with an evolving boundary,
to be determined by a model for an 
immiscible polymer brush-polymer melt interface and which 
in turn will be influenced by the flow.
This scope clearly demands extensive microscopic modeling
and we rather use a simple estimate, motivated by experiments. 
Interdiffusion dynamics for the symmetric case has been studied 
for both a melt and a network of PS on a PS-brush. 
The interface width has been found to change by a factor of 3-4 on the time scale
of a reptation time \cite{Clarke,Geoghegan} and displayed a logarithmic
behavior in time. For the immiscible system PS-PMMA, also a logarithmic growth 
in time has been found \cite{SferrazzaPML}, and has been explained by capillary wave theory.
For the purpose of demonstrating the influence on the dewetting,
we will assume a logarithmic time dependence of the roughness. 
In combination with $\zeta\propto \delta^2$, 
as a rough estimate for the time dependence of the friction coefficient 
we thus suggest the following expression: 
\beq\label{zeta_eq2}
\zeta(t)=\zeta_0+\zeta_1\log^2\left(1+t/t^*\right)\,.
\eeq
Here $\zeta_0$ is the friction coefficient 
associated to the status of the interface upon 
start of the dewetting process (by heating above the PS-glass temperature), 
$\zeta_1$ governs the importance
of the temporal evolution and $t^*$ is a short time scale
(the time $t^*$ has to be introduced on dimensional grounds; 
we assume $t^*\ll\tau_0,\tau_1$ as defined in Eq.~(\ref{const}) below).
We admit that the dependence given by Eq.~(\ref{zeta_eq2}) is empirical 
and a microscopic foundation would be desirable. 
However, as shown below, the effects of an increasing friction on dewetting 
turn out to be very robust with respect to variations of the detailed form of $\zeta(t)$.

\section{Modeling}

In order to demonstrate that a time evolution of interfacial properties
indeed can give rise to a maximum in the rim width and a concomitant dependence on sample age, 
we performed a numerical study for the model established recently in 
Refs.~\cite{Vilmin_nlfric,Raphael:2006.1,FZER1}.
This model is a lubrication approximation for a Jeffrey-type viscoelastic fluid film 
on slippery substrate. As motivated by recent experiments on polymer-polymer friction \cite{Leger},
nonlinear friction with the substrate is considered.
In the model, one accounts for three fields: the velocity in the film  $v(x,t)$, 
the stress field $\sigma(x,t)$, 
and the height profile $h(x,t)$ [assumed to be small compared to the slip length].
A simple edge geometry is considered, where the edge of the film 
is initially at $x=0$ and the dewetting occurs in $+x$-direction.
As governing equations one has the momentum equation (\ref{mech}), 
a Jeffrey-type 
constitutive law (\ref{const}), and volume conservation (\ref{heq})
\beq 
\label{mech}
\zeta \bar{v}^\al v^{1-\al}&=&\p_x\left(h\sigma\right)\,,\\
\label{const}
\sigma+\tau_1\p_t\sigma&=&G\tau_1\left[\left(\p_x v\right)+\tau_0\p_t\left(\p_x v\right)\right]\,,\\
\label{heq}
\p_{t} h&=&-\p_{x}\left(vh\right)\,.
\eeq
Eq.~(\ref{mech}) is the balance of the frictional force at the film-substrate interface
and the divergence of the total stress inside the film.
The exponent $\al\in[0,1)$ 
characterizes the nonlinear behavior of friction and is chosen to be
$\al=0.8$, a value obtained from recent experiments on high molecular weight PS
dewetting on PDMS-covered substrates \cite{Vilmin_nlfric} and 
consistent with measurements of rubber-brush friction \cite{Leger}. 
$\bar v$ is a characteristic velocity that enters the characteristic scales,
see Eq.~(\ref{scalea}) below. 
In Eq.~(\ref{const}), $G$ is the elastic modulus and $\tau_0$ and $\tau_1$, with $\tau_0\ll\tau_1$, 
are two characteristic time scales (with
$\tau_1$ of the order of the reptation time). 
These define a short-time viscosity $\eta_0=G\tau_0$
and a long-time viscosity $\eta_1=G\tau_1\gg\eta_0$.
As boundary conditions,
at the edge of the film the height-integrated stress has to equal the driving force,
$h(L)\sigma(L)=-\aS$. Additionally we impose $v(x$=$\infty)$=$0$, 
i.e. the film is unperturbed far away from the edge, and 
as initial conditions a quiescent film of thickness $h_0$.

Solving Eqs.~(\ref{mech}-\ref{heq}) analytically for short times, 
one can establish the characteristic velocity scale $V_{0,\al}$ and 
the characteristic width of the rim $W_{0,\al}$
\beq
\label{scalea}
V_{0,\al}=c\left(
\frac{V_0^2}{\bar{v}^\al}\right)^{\hspace{-1mm}\frac{1}{2-\al}},\,\,\,
W_{0,\al}=c\,W_0\left(
\frac{V_0^\al}{\bar{v}^\al}\right)^{\hspace{-1mm}\frac{1}{2-\al}}.
\eeq
Here $c=\left(\frac{2-\al}{2}\right)^{\frac{1}{2-\al}}$
is a numerical prefactor and
\beq
\label{scale}
V_0=\frac{|S|}{\sqrt{2\eta\zeta h_0}}\,\,\,,\,\,\,
W_0=\sqrt{\frac{\eta h_0}{\zeta}}\,.
\eeq
are the scales 
for a linear friction law; in that simple case, 
the characteristic width 
can be rewritten using the slippage length,  
$W_0=\sqrt{h_0 b}$, an expression already proposed some time ago
\cite{FBW:1994}. A rescaled version of the model 
($x'=x/\Woa$, $v'=v/\Voa$, $t'=t/\tau_0$) 
can then be solved
by a shooting method on a moving grid in space and Euler iteration in time.
For more details on the model and its solution we refer to \cite{Raphael:2006.1,FZER1}.
There, the model
has been studied in detail 
in the presence of a homogeneous initial residual stress 
(i.e. with initial condition  $\sigma(x,t$=$0)=\sigma_0\neq0$), 
that relaxes upon dewetting and thus giving rise to a maximum in the width. 
Assuming low molecular weight for the PS film, 
here we neglect the possibility of residual stress ($\sigma(x,t$=$0)=0$). 
Instead, motivated by the roughening of the 
interface found in \cite{Pascal_new} that should lead to increasing friction
with the substrate, 
in Eq.~(\ref{mech}) we implemented the time-dependent friction coefficient $\zeta(t)$ 
as given by Eq.~(\ref{zeta_eq2}). 

To test the robustness of the dewetting behavior
upon variations in the interface dynamics,
in addition we investigated a friction evolution like 
\beq\label{zeta_eq}
\zeta(t)=\zeta_0+(\zeta_{\infty}-\zeta_0)\left[1-\exp(-t/\tau^*)\right]\,.
\eeq
This dependence is motivated by the fact that
the roughness, and thus the friction coefficient, 
should saturate for long times, both because of the enthalpy cost given by $\chi$ and
due to the grafting of the PDMS chains. 

\begin{figure}[t!]
\vspace{5mm}
\begin{center}
	\includegraphics[width=.48\textwidth]{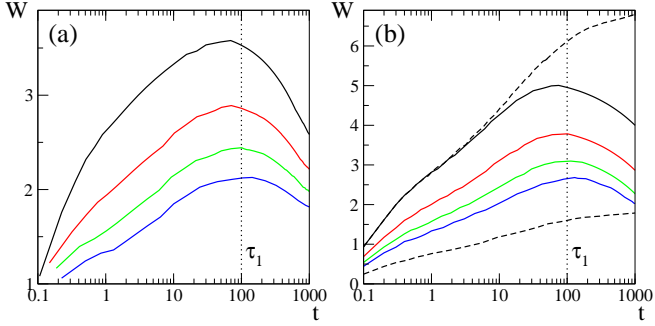}
	\caption{\label{figw}
	The width $W$ of the dewetting rim as a function of time in semi-logarithmic scale
	and for different initial values of the friction coefficient, $\zeta_0$.
	In panel (a) the friction coefficient evolves as given by Eq.~(\ref{zeta_eq2}) 
	[parameters $t^*=0.1$, $\zeta_1=0.03$] and in panel (b) by Eq.~(\ref{zeta_eq})
	[parameters $\tau_\zeta=500$, $\zeta_\infty=5$].
	Viscoelastic time scales $\tau_0=1$, $\tau_1=100$ and friction exponent $\alpha=0.8$.
	$\zeta_0=1$ (black), $1.5$ (red), $2$ (green), $2.5$ (blue).
	The dashed curves in panel (b) have been obtained with constant
	friction coefficients $\zeta=\zeta_0=1$ (upper curve) and 
	$\zeta=\zeta_\infty=5$ (lower curve).
	}
\end{center}
\end{figure}

\begin{figure}[t!]
\vspace{5mm}
\begin{center}
	\includegraphics[width=.48\textwidth]{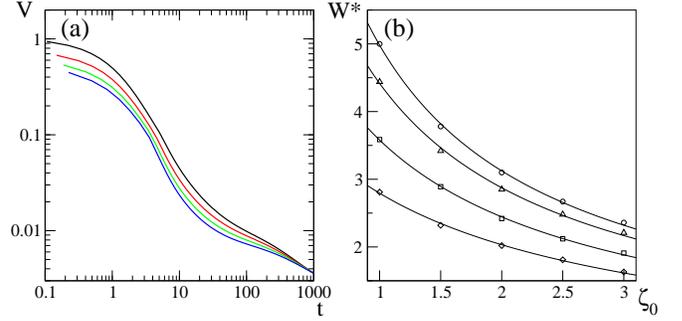}
	\caption{\label{figv}
	(a) The velocity $V$ at the edge in double logarithmic scale 
	for the friction evolution 
	given by Eq.~(\ref{zeta_eq})  
	[colors correspond to the initial values of 
	$\zeta_0$ as used in Fig.~\ref{figw}(b); same parameters].
	A rapid initial decay of the velocity is found, 
	only weakly depending on $\zeta_0$. 
	(b) The maximum of $W$, $W^*$, as a function of $\zeta_0$ for 
	the friction laws given by
	Eq.~(\ref{zeta_eq2}) [$\zeta_1=0.03$ ($\square$) , $\zeta_1=0.06$ ($\Diamond$); $t^*=0.1$] 
	and by Eq.~(\ref{zeta_eq})  
	[$\tau^*=500$ ($\circ$), 
	$\tau^*=200$  ($\vartriangle$); $\zeta_\infty=5$].
	Solid lines are fits to the estimate given by Eq.~(\ref{estimate}).
	}
\end{center}
\end{figure}

Numerical results for the temporal evolution of the width of the rim 
are shown in Fig.~\ref{figw}. 
One can see that the two different evolution laws for the friction coefficient, 
Eqs.~(\ref{zeta_eq2}) and (\ref{zeta_eq}), 
amount only to quantitative differences. 
The mechanism is robust and the qualitative behavior, (i) 
the occurrence of a maximum in the width and (ii) its monotonous decrease as a 
function of increasing $\zeta_0$, is unchanged.
For the friction law with saturation, Eq.~(\ref{zeta_eq}), it is instructive
to compare with the cases of non-evolving friction coefficients 
fixed at $\zeta_0$ and $\zeta_\infty$, see the upper and lower dashed curves, respectively, 
in Fig.~\ref{figw}(b). In both cases, there is no maximum
in the width. The two curves enframe the ones with evolving friction. The
curves with evolving friction will finally join for long times
with the curve obtained with $\zeta_\infty$ 
(however, the present model should not be applied for such long times).
Fig.~\ref{figv}(a) shows the dewetting velocity as a function of time.
A fast decay of the velocity in the elastic regime, for $\tau_0<t<\tau_1$, is obtained
as also observed experimentally. This signature is due to the decrease of 
the 'effective' driving force $|S|/\zeta$, and has also been observed in the model
when including 
a relaxing residual stress \cite{Raphael:2006.1,FZER1}.

Let us now discuss the effect of aging. Even if kept well below the bulk-PS
glass transition, the PS-PDMS interface evolves towards its equilibrium width,
as evidenced in \cite{Pascal_new}.
This is most probably due to the fact that motion at the interface is easier than in bulk 
and that, for the aging temperatures used, the PDMS is not glassy. 
Consequently this means that aging time translates
into increasing values of $\zeta_0$, i.e. of the value of the friction coefficient
at the moment where the dewetting process is initiated (experimentally by 
increasing the temperature above $T_g$).
As one can see from Fig.~\ref{figw}, increasing $\zeta_0$ leads to a 
{\it monotonous decrease} in the maximum of the width, 
in accordance with the experiments \cite{Pascal_new}.
Fig.~\ref{figv}(b) shows the dependence of the maximum in the width as
a function of the initial value of the friction coefficient, $\zeta_0$.
There is consistent decrease for both proposed evolutions for the friction coefficient 
and for various parameter values.

The numerically obtained results can also be interpreted
by a scaling estimate: the typical scale for the characteristic width, Eqs.~(\ref{scalea},\ref{scale}),
implies a dependence on viscosity $\eta$ and friction $\zeta$ as
\beq\label{estimate}
W\propto\eta^{\frac{1-\alpha}{2-\alpha}}\zeta^{-\frac{1}{2-\alpha}}\,,
\eeq 
where $\alpha$ is the nonlinear friction exponent.
At short times, one has to use $\eta_0$ and $\zeta_0$ in this formula,
while at times $t>\tau_1$ one has $\eta_1$ and $\zeta(t)$.
This provides a simple picture for the occurrence of the maximum:
The increase of the viscosity leads to an increase in the width of the rim
up to a time of the order of $\tau_1$. Later on, the width will decrease slowly
since the viscosity stays at $\eta_1$, while the friction coefficient $\zeta$ 
still slowly increases.
Assuming the maximum $W^*$ to be roughly at $\tau_1$, cf. Fig.~\ref{figw}, 
one can estimate 
$W^*\propto\left[\zeta(\tau_1)\right]^{-\frac{1}{2-\alpha}}$.
This expression was used to fit the dependence of $W^*$ as a function
of $\zeta_0$ to obtain the solid lines in Fig.~\ref{figv}(b), with
a fixed value $\alpha=0.8$ for the exponent, as used in the numerical
solution.
Although the window of $\zeta_0$-values is too small to
establish a real power law, the agreement of this simple estimate 
is very good.

The numerical study and the simple scaling thus suggest that
- once the temporal increase in friction due to roughening
is established by a microscopic theory -
it would be possible to measure the roughness evolution below the glass temperature,
i.e. the aging of the buried polymer-polymer interface,
by tracing the maximum in the rim width in a dewetting experiment. 

We should mention that
in case of the interface evolution studied here, 
a maximum in the width of the rim 
can already be obtained for a linear friction law, i.e. for $\alpha=0$
in the momentum equation (\ref{mech}).
This is in contrast to the case where relaxing residual stresses
are responsible for the maximum, as discussed in detail in Refs.~\cite{Raphael:2006.1,FZER1},
where nonlinear friction was crucial.
The underlying reason for this difference is the fact 
that the friction coefficient already influences
$W$ for a linear friction law, see Eq.~(\ref{estimate}) with $\alpha=0$,
while for the case with residual stress the nonlinearity in the
friction was needed to introduce a dependence on residual stress in the scaling for $W$.
However, looking at the scaling law Eq.~(\ref{estimate}), 
for $\alpha=0$ the dependence on the friction coefficient is
much weaker, $W\propto\zeta^{-1/2}$, as compared to $W\propto\zeta^{-5/6}$ for $\alpha=0.8$. 
Hence a much larger increase in friction is needed if the friction law is linear, 
which might be improbable
for the small interface changes that are to be expected
for an immiscible polymer-polymer interface. 
Thus, although nonlinear friction is not essential
it amplifies the effect of the interface evolution with respect to
the viscosity evolution due to viscoelasticity, as directly shown by Eq.~(\ref{estimate}). 
Investigating the (non-)linearity of the friction experimentally, similarly 
as has been done in Ref.~\cite{Vilmin_nlfric},
might thus be helpful to better understand quantitative features of the 
maximum in the rim width.

\section{Conclusions and perspective}

To conclude, an increase in friction (or a decrease in the slip length) 
due to an evolution of the 
substrate-film 
interface 
can give rise to a maximum in the rim width upon dewetting.
This implies that the  - macroscopically observable - maximum in the width 
may be caused by and is susceptible to minute changes of the interface. 
In addition, there is a monotonous relation between the maximum of the width 
and the friction coefficient at the beginning of dewetting. 
Concerning the experiments on PS-PDMS systems, 
this friction coefficient in turn is affected by the aging time 
of the polymer-polymer interface below the glass temperature.
As compared to the relaxation of residual stresses put forward recently 
\cite{Reiter:2005,Vilmin_nlfric,Raphael:2006.1},
our work offers an alternative interpretation of these aging effects, 
in the case of low-molecular weight dewetting films.
However, at present stage it is not possible to give a direct relation 
between the maximum in the rim width and the aging time, 
because neither the exact time evolution of the interface towards its equilibrium value, 
nor the exact relation (for an asymmetric melt-brush system) 
between interface width/roughness and friction/slippage is known.
We would like to encourage MD simulations or dynamic self-consistent field studies to
establish this dependence. Then it would be possible to investigate
buried polymer-polymer interfaces by such relatively simple and 
cheap experiments as dewetting.

\acknowledgments

We thank Pascal Damman for inspiring this work and Liliane L\'eger 
and G\"unter Reiter for very stimulating discussions.
F.Z. acknowledges financial support by the German Science Foundation (DFG).

\end{document}